\documentclass[aps,prl,twocolumn,superscriptaddress,showpacs]{revtex4}
\usepackage{graphicx}
\usepackage{multirow}
\usepackage{amsfonts}
\usepackage{amssymb}
\usepackage{amsmath}

\def\cm{cm$^{-1}$}
\usepackage{epstopdf}

\begin{document}
\title{Boson peak in overdoped manganites La$_{1-x}$Ca$_x$MnO$_3$}
\author{B.~Gorshunov}
\affiliation{1. Physikalisches Institut, Universit{\"a}t Stuttgart, ~Pfaffenwaldring 57, 70550 Stuttgart, Germany}
\affiliation{A.M. Prokhorov General Physics Institute, Russian Academy of Sciences,
%Vavilov str.\ 38,
119991 Moscow, Russia}
\affiliation{Moscow Institute of Physics and Technology (State University), 141700, Dolgoprudny, Moscow Region, Russia}
\author{E.~Zhukova}
\affiliation{1. Physikalisches Institut, Universit{\"a}t Stuttgart, ~Pfaffenwaldring 57, 70550 Stuttgart, Germany}
\affiliation{A.M. Prokhorov General Physics Institute, Russian Academy of Sciences,
%Vavilov str.\ 38,
119991 Moscow, Russia}
\affiliation{Moscow Institute of Physics and Technology (State University), 141700, Dolgoprudny, Moscow Region, Russia}
\author{V.I.~Torgashev}
\affiliation{Faculty of Physics, Southern Federal University, %Zorge Str. 5,
344090 Rostov-on-Don, Russia}
\author{L.S.~Kadyrov}
\affiliation{A.M. Prokhorov General Physics Institute, Russian Academy of Sciences,
%Vavilov str.\ 38,
119991 Moscow, Russia}
\affiliation{Moscow Institute of Physics and Technology (State University), 141700, Dolgoprudny, Moscow Region, Russia}
\author{L.~Motovilova}
\affiliation{1. Physikalisches Institut, Universit{\"a}t Stuttgart, ~Pfaffenwaldring 57, 70550 Stuttgart, Germany}
\affiliation{A.M. Prokhorov General Physics Institute, Russian Academy of Sciences,
%Vavilov str.\ 38,
119991 Moscow, Russia}
\affiliation{Moscow Institute of Physics and Technology (State University), 141700, Dolgoprudny, Moscow Region, Russia}
\author{F.~Fischgrabe}
\affiliation{I.~Physikalisches Institut, Georg-August-Universit\"at G\"ottingen,
%Friedrich Hund Platz,
37077 G\"ottingen, Germany}
\author{V.~Moshnyaga}
\affiliation{I.~Physikalisches Institut, Georg-August-Universit\"at G\"ottingen,
%Friedrich Hund Platz,
37077 G\"ottingen, Germany}
\author{T.~Zhang}
\affiliation{Key Laboratory of Materials Physics, Institute of Solid State Physics, Chinese Academy of Sciences, Hefei 230031, People's Republic of China}
\author{R.~Kremer}
\affiliation{Max-Planck-Institut f\"ur Festk\"orperforschung, %Heisenbergstrasse 1,
70569  Stuttgart, Germany}
\author{U.~Pracht}
\affiliation{1. Physikalisches Institut, Universit{\"a}t Stuttgart, ~Pfaffenwaldring 57, 70550 Stuttgart, Germany}
\author{S. Zapf}
\affiliation{1. Physikalisches Institut, Universit{\"a}t Stuttgart, ~Pfaffenwaldring 57, 70550 Stuttgart, Germany}
\author{M.~Dressel}
\affiliation{1. Physikalisches Institut, Universit{\"a}t Stuttgart, ~Pfaffenwaldring 57, 70550 Stuttgart, Germany}
\date{\today}

\begin{abstract}
In the charge-ordered phase of strongly doped manganites La$_{1-x}$Ca$_x$MnO$_3$ ($x \geq 0.5$) absorption lines appear in the terahertz spectral range for commensurate $x$ values right below the charge-ordering temperature. They are connected to acoustic phonons that become optically active by folding of the Brillouin zone. At  lower temperatures a strongly asymmetric extra absorption band develops at frequencies corresponding to the position of the lowest-energy van Hove singularity in the reduced Brillouin zone. The band is assigned to the boson peak, {\it i.e.}\ to the excess of lattice vibrational states over the standard Debye contribution. The folded phonons and the boson peak do not show up for incommensurate calcium contents when no distinct Brillouin zone folding exists.
\end{abstract}
\pacs{
63.50.-x, 	%Vibrational states in disordered systems
73.20.Mf,    %Collective excitations (including excitons, polarons, plasmons and other charge-density excitations)
71.45.Lr,    %Charge-density-wave systems (see also 75.30.Fv Spin-density waves)
75.47.Lx,    %Magnetic oxides}
75.25.Dk    %Orbital, charge, and other orders, including coupling of these orders
%78.30.-j,    %Infrared and Raman spectra
}
\maketitle

%\section{Introduction}
Manganites of the family  $R_{1-x}A_x$MnO$_3$ ($R$ is a rare earth and $A$ an alkaline element) provide a unique laboratory to study the physical properties of solids with several competing (or cooperating) order parameters.
It is of general importance to understand the transitions to the various single- or multi-phase ground states
because eventually it can provide ideas for solving fundamental problems in the physics of electronically correlated systems. The task, however, is severely complicated by the diversity of interactions of comparable strength in charge, magnetic, lattice or orbital sectors -- each trying to stabilize the system according to the corresponding order patterns. For instance, even the well-known colossal negative magnetoresistance effect in hole doped manganites is not understood in all details at this point \cite{Ramirez97,Dagotto03}.

Electronically doped (and in particular overdoped)
manganites are significantly less studied because they do not exhibit
the colossal magnetoresistance effect; instead they demonstrate a dielectric charge-ordered (CO)
ground state  that can coexist with antiferromagnetic order \cite{Salamon01,Chen96}.
The origin of the corresponding phase transition
has attracted increasing attention in recent years.
One of the most popular approaches is based on the charge-density wave (CDW) scenario \cite{Milward05}
because the systems seem to possess the necessary ingredients like the Fermi surface
nesting and lattice superstructure \cite{Tonouchi07,Chuang01}.
Low-energy excitations in the THz frequency range are frequently taken as evidence in favor
of this mechanism as they are ascribed to the CDW-condensate phase or amplitude modes \cite{Kida02,Nucara08,Mavani08,Fujioka10,Rana13}.
There is, however, some dispute about this Peierls-Fr{\"o}hlich scenario \cite{Schmidt08,Fisher10}. In particular,
the THz resonances can be self-consistently interpreted as acoustical lattice
vibrations that acquire optical activity when the Brillouin zone is folded
due to the superstructure in the CO state \cite{Pissas03,Zhang10}.
This superstructure and corresponding folded phonons are inherent fingerprints of the ordering in manganites. Hence the aim of the present work is to analyze the mechanisms of corresponding phase transitions by detailed exploration of the folded phonons in La$_{1-x}$Ca$_x$MnO$_3$ dependent on the calcium content in the broad interval $0.5 \leq x\leq 1$. An important motivation was also to clarify the role of disorder in the physics of overdoped La$_{1-x}$Ca$_x$MnO$_3$ manganites that are available mainly as polycrystals. It is well known that disorder can drastically change the low-energy phonon density of states (DOS) of a solid and can even lead to appearance of rather sharp absorption resonance -- the so called boson peak \cite{Phillips81,Elliott90}.
Similar effects are expected in La$_{1-x}$Ca$_x$MnO$_3$ for {\em incommensurate} calcium contents $x$ that leads to a certain degree of disorder in the lattice ions positions.
To which extent the disorder-driven renormalization of the phonon DOS
can be responsible for the low-energy physics of manganites and in particular for the soft excitations reported in the literature \cite{Galeener78,Kirk88,Taraskin01,Schirmacher98,Kida02,Rana13,Nucara08,Mavani08,Fujioka10}, remains unclear.

\begin{figure}%[!http]
\centering
\includegraphics[width=1\columnwidth]{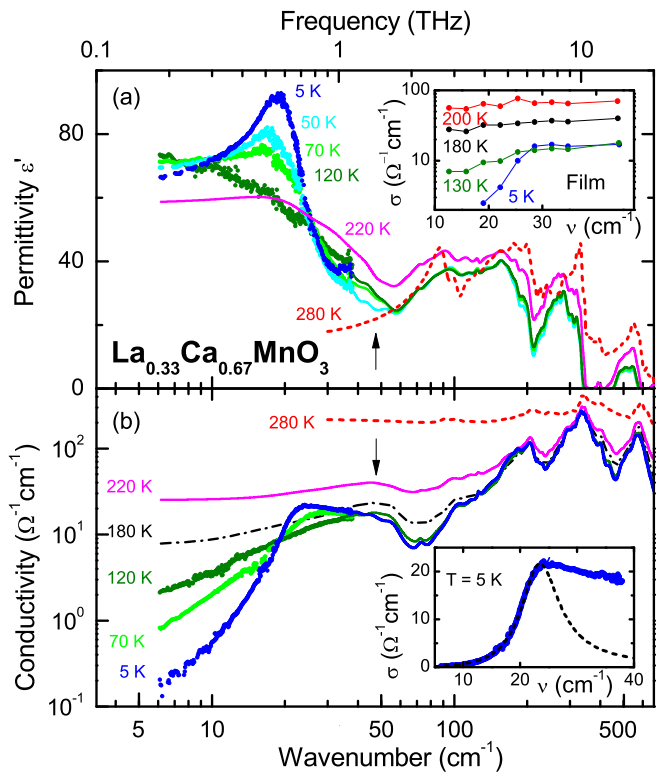}
\caption{(Color online) Temperature dependence of the optical properties of La$_{0.33}$Ca$_{0.67}$MnO$_3$ ceramic sample  in the THz range. In the low-frequency part of (a) the dielectric permittivity $\epsilon^{\prime}(\nu)$ and (b)~the optical conductivity $\sigma(\nu)$ the charge order transition at $T_{\rm CO}=240$~K is seen as dramatic change due to the opening of the energy gap in the far infrared. The arrows indicate the vibrational acoustic modes at 50~\cm\ that become infrared active by Brillouin zone folding due to charge ordering. The inset in panel (a) exhibits the temperature evolution of the THz conductivity measured on a thin epitaxial La$_{0.33}$Ca$_{0.67}$MnO$_3$ film. In the lower panel the inset demonstrates that it is not possible to fit the low-frequency asymmetrical band by a Lorentzian curve (dashed line).}
\label{fig:1}
\end{figure}
Polycrystalline samples La$_{1-x}$Ca$_x$MnO$_3$ (LCMO) with $x({\rm Ca})=0.5$; 0.6; $2/3\approx 0.67$; 0.7; 0.75; 0.85; 0.9; 0.95; 0.98 and $x=1$
 were prepared by the solid-state reaction method. Stoichiometric amounts of La$_2$O$_3$, CaCO$_3$, and MnO were mixed and ground together. The powders were calcined at 1100$^{\circ}$C in air for 12~h, reground, and reacted at 1200$^{\circ}$C  for another 12~h. These last steps were repeated three times followed by regrinding and reacting the powders at 1250$^{\circ}$C  for 12~h. The resultant materials were pressed into pellets of 2~mm thickness and subsequently sintered again at 1250$^{\circ}$C for 30~h in air. The specimens were slowly cooled to room temperature at a rate of 1$^{\circ}$C/min. All pellets were prepared in identical conditions. The Ca doping was additionally
controlled by x-ray diffraction measurements. In addition, epitaxial films were deposited on MgO substrates in a way described in Ref.~\onlinecite{Moshnyaga99}.  The films show a
strain-free state as indicated by close proximity of $c$-axis lattice
parameters of films and bulk samples.

For optical measurements two kinds of spectrometers were employed: In the range of THz frequencies (3 up to 40~\cm) we utilized a Mach-Zehnder interferometer based on backward-wave oscillators as coherent radiation sources \cite{Kozlov98,Gorshunov05}. The permittivity  $\epsilon^{\prime}(\nu)$ and the conductivity $\sigma(\nu$) spectra were directly determined from the complex transmission coefficient measured through specimens of thickness $d$ between $30~\mu$m and 1~mm. In addition a Fourier-transform spectrometer was utilized to measure the far-infrared reflection coefficient of the same samples, up to $\nu=\omega/(2\pi c)=700$~\cm. Kramers-Kronig and dispersion analysis of the reflectivity data in combination with the directly measured THz optical constants allowed us to obtain $\epsilon^{\prime}(\nu)$ and $\sigma(\nu)$  in the range approximately 3 to 700~\cm\ and at temperatures from 5 to 300~K.
The main attention of the present report is focused on excitations detected at low temperatures in the THz frequency range for all samples with $0.5\leq x < 0.85$.

Fig.~\ref{fig:1} presents the temperature evolution of the THz-infrared spectra of $\epsilon^{\prime}(\nu)$ and $\sigma(\nu)$ of LCMO with a commensurate
calcium content, $x=2/3$. In the sub-phonon frequency range, below 100~\cm,
unexpected modifications of the spectra occur in the CO state.
Right below  $T_{\rm CO}= 240$~K --~when the background conductivity of the free-carriers freezes out~-- a bump becomes visible in $\sigma(\nu)$ at around 50~\cm\  [Fig.~\ref{fig:1}(b)];
at the lowest temperatures at least two separate peaks can be distinguished.
Below approximately 120~K an additional strong absorption band develops at slightly lower frequencies between 20 and 30~\cm; note, the intensity (spectral weight) in this spectral range grows significantly as the temperature decreases. In the linear presentation
of $\sigma(\nu)$ [inset of Fig.~\ref{fig:1}(b)] the asymmetric lineshape of the band becomes
obvious;  it is not possible to fit it by a Lorentzian.
Related to this asymmetric band the permittivity spectra exhibits an unusual
dispersion [Fig.~\ref{fig:1}(a)] that becomes dominant
when the temperature is reduced to 5~K. Note that the $\epsilon^{\prime}(\nu)$ spectrum at $T=5$~K demonstrates underdamped character of the  absorption band, {\it i.e.} the permittivity first increases with frequency and then drops in the vicinity of the resonance frequency.

\begin{figure}
\centering
\includegraphics[width=0.8\columnwidth]{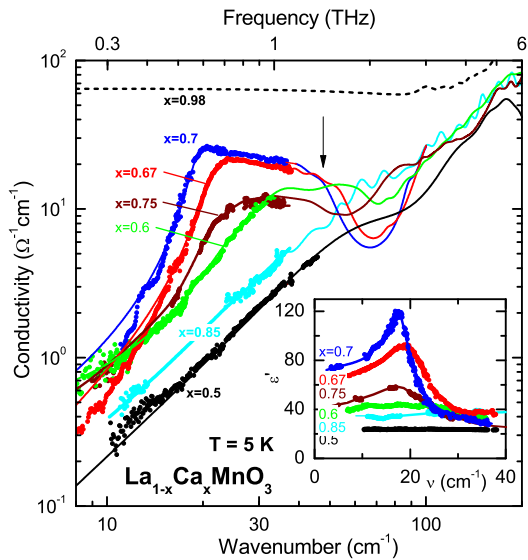}
\caption{(Color online) Development of the low-temperature terahertz conductivity of La$_{1-x}$Ca$_{x}$MnO$_3$ ceramics for different $x$. Close to $x=2/3$ an asymmetric  band develops around $20-30$~\cm,
that disappears again upon further doping.
The charge-order activated folded modes are indicated by the arrow. Spectrum for $x=0.75$ was taken on a pellet pressed of a powder \cite{Zhang10}. The dielectric permittivity $\epsilon^{\prime}(\nu)$ is displayed in the inset for the same compositions.}
\label{fig:2}
\end{figure}

In Fig.~\ref{fig:2} the development of the THz absorption bands is presented for the series of compounds with calcium contents $x=0.5$, 0.6, 2/3, 0.7, 0.85, 0.98. It is important to note that in all samples that reveal transition to the CO phase ($0.5< x\leq 0.75$) the 50~\cm\ absorption start to appear right below $T_{\rm CO}$. For commensurate ($x=2/3$) or almost commensurate ($x=0.7$) calcium contents the strong lower-frequency resonance is clearly present. It always exhibits the pronounced asymmetric shape as discussed above [inset of Fig.~\ref{fig:1}(b)]].

Similar asymmetric absorption bands in LCMO were observed previously \cite{Kida02,Nucara08,Mavani08,Fujioka10,Rana13}
and assigned to collective excitations of the CDW ground state.
However, there are good arguments against this interpretation: (i)~The bands consist of two or even more spectral components that could only be explained by multi-phonon Peierls coupling, as in the case of organic mixed stack compounds \cite{Girlando08}. (ii)~The pinned CDW modes are typically heavily overdamped \cite{Gruner94,Cox08}, in contrast to our observations [cf.\ underdamped character of dispersion of permittivity spectra in Fig.~\ref{fig:1}(a) and in inset in Fig.~\ref{fig:2}]. (iii)~One expects different frequencies of the pinned modes when varying  the doping values $x$ \cite{Degiorgi91}. (iv)~The dielectric contribution $\Delta\epsilon$ of the CDW pinned modes can reach very high values (up to $10^{4} - 10^{6}$ \cite{Gruner94}), much larger than in the present case, where $\Delta\epsilon \approx 30-40$ (inset of Fig.~\ref{fig:2}). (v)~The ratio $2\Delta/k_BT_{\rm CO}$ estimated from the optical gap $2\Delta$ is too large (approximately 6 to 30 \cite{Kim02,Zhang10}) when compared to the mean-field value of 3.5.

\begin{figure}
\centering
\includegraphics[width=0.8\columnwidth]{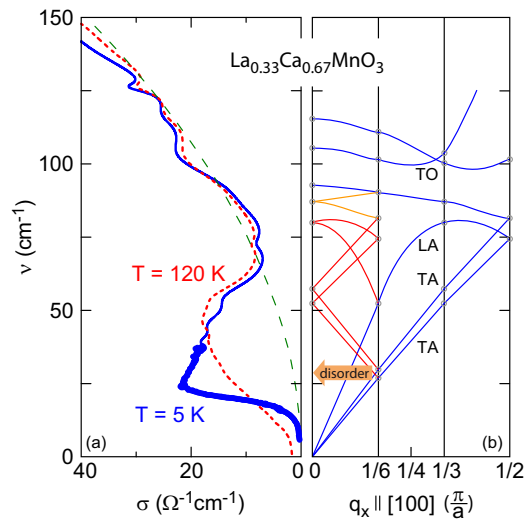}
\caption{(Color online) Explanation of the observed conductivity spectra (that is proportional to absorption) by the vibrational density of states. (a)~Optical conductivity of La$_{0.33}$Ca$_{0.67}$MnO$_3$ at two different temperatures. The dashed line corresponds to a $\sigma(\nu)\propto \nu^2$ behavior. (b)~Energy dispersion of lattice vibrations calculated for CaMnO$_3$. For commensurate doping $x=2/3$ the Brillouin zone is folded to one third and the acoustic phonons become optical active around 50~\cm. In the case of disorder as in polycrystalline samples, a broad distribution with a maximum in the density of states around 25~\cm\ occurs, called the boson peak.}
\label{fig:3}
\end{figure}
We associate the observed low-energy absorption bands with acoustical lattice vibrations that become infrared activated due to the Brillouin zone folding. The CO formation in manganites is accompanied by a lattice superstructure along the $a$-direction with a period $a^*=a(1-x)$, $a$ is the lattice constant \cite{Larochelle01,Loudon05}. As demonstrated in Fig.~\ref{fig:3}(b) by a sketch of a Brillouin zone for $x=2/3$ compound, due to the triple folding of the Brillouin zone additional (folded) phonon modes should appear at around 50~\cm. In fact, they can be resolved in the spectra  (Fig.~\ref{fig:1}) as weak peaks on the high-frequency shoulder of the pronounced 25~\cm\ band.
This strong latter band, however, cannot be explained by a regular folded phonon, because it is located below 50~\cm\ where no folded phonons are expected and most important, because it exhibits a remarkably asymmetric lineshape. We argue that this asymmetric band represents the DOS of acoustical phonons that appears in the spectra due to the disordered nature of our polycrystalline samples.

The disorder-induced reconstruction of the low-energy part of the vibrational DOS $G(\nu)$ is known from glasses, amorphous materials and similar disordered crystalline solids. It results in extra states, in addition to the Debye part that is proportional to the squared frequency $\nu^2$, and is experimentally seen in the THz frequency range in the form of a peak in the reduced DOS $g(\nu)=G(\nu)/\nu^2$. This so-called boson peak \cite{Phillips81,Elliott90} is one of the most discussed topics and most puzzling open questions in disordered systems. Numerous models are formulated  to explain the mechanisms of the boson peak (see \cite{Schirmacher98,Taraskin99,Gurevich03,Chumakov11} and references therein) but its origin is still not completely clear. Inelastic neutron scattering might be the best probe to observe the boson peak, nevertheless due to the disorder-driven breakdown of the selection rules it is also seen in optical spectra \cite{Galeener78,Kirk88,Taraskin01,Schirmacher98}. Based on numerical calculations \cite{Taraskin01} and experimental observations \cite{Chumakov11} the boson peak is associated with the critical points within the Brillouin zone, more specifically,  with the lowest-energy van Hove singularity of the crystal.
%It commonly shows up at rather low frequencies when disorder is introduced.

The described signatures of the boson peak are in remarkable qualitative and even quantitative agreement with our data on La$_{1-x}$Ca$_x$MnO$_3$ where the polycrystalline nature of  the samples should introduce a certain degree of disorder, such as the random distribution of force constants and/or charge fluctuations at the grain boundaries. Indeed, the resonances in the spectra of LCMO samples with commensurate $x=2/3$ and almost commensurate $x=0.7$ concentrations emerge around $20-25$~\cm, that is slightly below the lowest-energy van Hove singularities of the three-fold reduced Brillouin zone (Figs.~\ref{fig:1}, \ref{fig:2} and \ref{fig:3}), in full agreement with the theoretical prediction \cite{Taraskin01}. It should be pointed out that the asymmetric lineshapes of these resonances closely resemble typical shapes of the
vibrational DOS near the Brillouin zone  critical points.
It is interesting to compare the spectra of ceramic sample $x=2/3$ with those of epitaxial films of the same composition, where the lower degree of disorder leads to a significantly reduced strength of the boson peak, as can be seen in the inset of Fig.~\ref{fig:1}(a).

A boson peak in the phonon DOS may also become obvious in an additional term of the low-temperature lattice contribution to the heat capacity, e.g.\ by a deviation from Debye's $T^3$ power law. According to the general trend that noticeable contributions from a peak in the phonon DOS become apparent at temperature of $\sim 1/6$ of the peak energy \cite{Cardona10}, we expect for La$_{0.33}$Ca$_{0.67}$MnO$_{3}$ deviations from the Debye law below  5~K. Our experiments on La$_{0.33}$Ca$_{0.67}$MnO$_{3}$ down to 2~K indeed revealed additional low-temperature heat capacity contributions. At present, however, it is not clear whether
additional magnetic contributions have to be considered besides the extra contribution
of the boson peak.

When turning to commensurate doping $x=1/2$, we can understand from Fig.~\ref{fig:3} that the phonon-related low-frequency features should be shifted to higher frequencies compared to the $x=2/3$ compound, since the superlattice period is doubled (not tripled as for $x=2/3$), {\it i.e.} the Brillouin zone should be reduced to $\pi/4a$ compared to $\pi/6a$ . From Fig.~\ref{fig:2} we see, however, that these features can hardly be resolved.
The reason is that the composition $x=1/2$ falls right on the boundary between the conducting ferromagnetic ($x<0.5$) and insulating CO ($x>0.5$) phases. The multi-phase character of this compound masks the well-formed crystal lattice superstructure and blurs the Brillouin zone  picture.
In the case of the {\em incommensurate} value $x=0.6$ the THz absorption band is quite pronounced and also asymmetric. This calcium content might be seen as a mixture of neighboring commensurate phases, $x=1/2$ and $2/3$ \cite{Pissas03}; accordingly the observed band is composed of folded phonons and of signatures of the boson peaks from these phases. For even higher doping level, $x=0.85$, the THz spectra do not contain any structures because no CO develops and thus no superlattice exists. The same holds for compound with $x=0.98$ where, in addition, the high conductivity level screens out the THz structures.

\begin{figure}
\centering
\includegraphics[width=0.6\columnwidth]{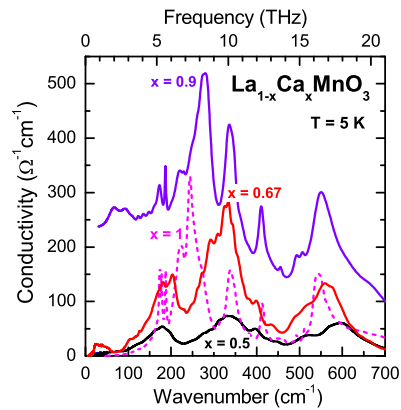}
\caption{(Color online)Phonon modes of La$_{1-x}$Ca$_x$MnO$_3$ for different $x=0.5$, $2/3$, 0.9 and 1. While for $x<0.87$ three broad bands are observed that shift to lower frequencies as $x$ increases, above $x=0.87$ the absorption features change to a typical phonon spectrum.}
\label{fig:4}
\end{figure}
In Fig.~\ref{fig:4} the far-infrared spectra of some selected
La$_{1-x}$Ca$_x$MnO$_3$ samples are presented on a linear  scale in order to elucidate the phonons features. It is noticeable  that the above-discussed acoustics-related excitations (folded phonons and boson peaks) are of significantly smaller amplitudes compared to the first-order phonon  bands. This is an indication that charge order induces only a relatively small weight fraction of lattice modulations \cite{Pissas03}.
As for the first order phonon lines in La$_{1-x}$Ca$_x$MnO$_3$ that are presented in Fig.~\ref{fig:4}, the spectral density of phonon states for compounds with $0.5 \leq x < 0.87$ is composed by three continua that correspond to the transverse optical (TO) cubic perovskite modes TO1, TO2 and
TO3. For commensurate modulations of the lattice (like for $x = 2/3$) the phonon resonances are well structured that
is a consequence of transition to a long-periodic structure in the CO state. The phonon spectra change drastically
when the concentrational phase transition at $x = 0.87$ is crossed.
The typical phonon response for higher concentrations (represented in Fig.~\ref{fig:4} by the $x = 0.9$ spectrum)  demonstrates a higher degree of order in the crystal lattice as compared to the $x < 0.87$ solid
solutions. The phonon DOS for CaMnO$_3$ is typical  for the rhombic perovskites of the
$Pnma$ symmetry with four molecules in the unit cell.

In conclusion, absorption bands with complex structures are discovered  below $T_{\rm CO}$ in the terahertz spectra of overdoped manganites La$_{1-x}$Ca$_x$MnO$_3$ for commensurate calcium concentration. We identify acoustical phonon modes which are optically  activated in the charge ordered state where the crystal lattice
forms a superstructure and the corresponding Brillouin zone is folded back. Most important, we observe a pronounced extra absorption feature with strongly asymmetric lineshape, that we associate  with the boson peak -- a counterpart of the acoustic van Hove singularity at the boundary of the folded Brillouin zone that gains optical activity due to disorder.
Our findings evidence that disorder effects must be taken into account when analyzing low-energy properties of overdoped manganites.

We acknowledge fnancial support by the Deutsche Forschungsgemeinschaft (DFG) via DR228/36, Russian State contracts N2011-1.2.1-121-003, 14.18.21.0740,
RFBR project 11-02-91340-HHNO-a, and by the Natural Science Foundation of China  Grant No. 10904146). We enjoyed discussions with S.N. Taraskin.

\end{document}